\documentclass[prl,aps,letterpaper,floatfix,superscriptaddress,preprintnumbers,twocolumn,10pt]{revtex4-1}

\usepackage{dcolumn}
\usepackage{bm}
\usepackage[dvips]{graphicx}
\usepackage{epsfig}
\usepackage{amsmath, amsfonts, amssymb, slashed}

\newcommand{\be}{\begin{equation}}
\newcommand{\ee}{\end{equation}}
\newcommand{\bee}{\begin{equation*}}
\newcommand{\eee}{\end{equation*}}

\usepackage{color}

\newcommand{\bea}{\begin{eqnarray}}  
\newcommand{\eea}{\end{eqnarray}}

\begin{document}

\title{Theoretical interpretation of a spin-two diphoton excess}

\author{Ver\'onica Sanz} 

\affiliation{Department of Physics and Astronomy, University of Sussex, 
Brighton BN1 9QH, UK}

\date{\today}

\begin{abstract}
A spin-two resonance would have couplings to the SM, suppressed via higher-dimensional operators. Therefore, a sizeable cross-section would indicate that the scale of additional new phenomena cannot be too far above the 750 GeV mark. Below I present some theoretical arguments to place the resonance in a wider framework: a manifestation of new extra-dimensions, or a glueball of a new strong sector. In each case, I identify some conditions the UV theory must satisfy to accommodate other experimental bounds. Based on those, I draw correlations with associated phenomena we expect to find in future searches.  
\end{abstract}

\maketitle
\section{Introduction}

As I write this note, ATLAS and CMS collaborations are presenting their results on the diphoton excess at Moriond, an update from the December results~\cite{December}. Both collaborations re-analyzed their data in terms of two hypothesis, spin-zero and -two new resonances. ATLAS obtains a local significance of about 3.5$\sigma$ in both cases, and explores the compatibility of the data with the Run1. CMS combines Run1 and Run2, finding a local significance of about 3.4$\sigma$ in both cases. Finally, both experiments quote a global significance in the range of 1 to 2$\sigma$, although these numbers neglect the important fact that both collaborations see the excess in the {\it same} region. Obviously, we should be cautious and wait for more data to increase (or decrease) the global significance of this excess, most likely during the late summer conferences. 

In the meantime we are going to entertain a few ideas as to the origin of this excess. Any hint for new physics is exciting news for us, but a spin-two resonance would be particularly surprising, as it is not the type of new physics most people would consider at first. To help in understanding this result from a theoretical perspective, I am going to place this resonance into a wider perspective, and provide some guidance on other phenomena we would then expect to observe.

\section{General considerations\label{general}}
 
 Let us start by setting a model-independent parametrization of a spin-two resonance, which I denote by $X_{\mu\nu}$. The propagation of $X$ is described by the  Fierz-Pauli Lagrangian~\cite{Fierz:1939ix}. It describes a massive spin-two field via a rank-two symmetric and traceless tensor, i.e.$X_{\mu\nu}=X_{\nu\mu}$ and $X_{\mu}^{\mu} =0$,
which satisfies a positive-energy condition $\partial_{\mu} X^{\mu\nu} = 0$.
The form of the kinetic term is not particularly revealing, and can be found elsewhere, e.g. ~\cite{Hagiwara:2008jb}.

We also know that the resonance interacts. The interaction is likely to conserve Lorentz invariance, as one can see a bump in the diphoton invariant mass, which stems from momentum conservation in the decay. Asssuming then Lorentz invariance is preserved by $X$, interactions cannot come from a renormalizable term in the Lagrangian. Indeed, the lowest order one can write an $X$-$SM$-$SM$ interaction term is via dimension-five operators. With the further assumption that the resonance interactions preserve CP, one can for example write terms involving photons and gluons such as
\bea
{\cal L}_{int} \supset -\frac{c_\gamma}{\Lambda} \, X_{\mu\nu} \, F^{\mu\lambda} F^{\nu\lambda} -\frac{c_g}{\Lambda} \, X_{\mu\nu} \, G^{\mu\lambda} G^{\nu\lambda}  \ . 
\eea

One could also consider a CP odd spin-two case, but for simplicity I will continue the discussion with the CP even option. Note that there would be strong constraints from couplings of the resonance to light fermions in the CP-odd case~\cite{Fok:2012zk}. At any rate, the conclusion that this resonance interacts via higher-dimensional operators is valid in both the CP-odd and -even cases.

The upshot of this discussion is that $X$ must interact via non-renormalizable operators, characterized by a scale of new phenomena $\Lambda$. And, although $X$'s interactions are power suppressed, the signal rate in diphotons is relatively strong. This leads to the interesting conclusion that $\Lambda$ cannot be too far from the TeV scale. Indeed, this can be understood by looking at the gluon fusion rate to diphoton at LHC13~\cite{Han:2015cty},
\bea
\sigma(pp\rightarrow X\rightarrow \gamma\gamma)= 8.6 \textrm{pb}  \, (c_{g} c_{\gamma})^2\left(\frac{3 \textrm{TeV}}{\Lambda}\right)^4  \, \left(\frac{\textrm{GeV}}{\Gamma_X} \right) \ , \nonumber \\
\eea
where $\Gamma_X$ is the width of the resonance. 

So far this discussion has been kept in a model-independent fashion, but we expect new phenomena not far from the mass of $X$. To understand the origin of the interactions, and draw new predictions, we need to plunge into possible UV completions of the resonance.

\section{A Kaluza-Klein graviton? \label{KKG}}
 
An obvious candidate for a massive spin-two resonance near the TeV scale is a Kaluza-Klein (KK) graviton. This is the interpretation used by the experimental collaborations. 

If this were to be correct, it would imply the discovery of new dimensions of space. Indeed, a massive KK graviton is the {\it smoking gun} of extra-dimensions, or isn't it? In fact, it has been shown that Lorentz and CP invariance ensures that {\it any} spin-two resonance exhibits the same coupling to two SM particles as a KK-graviton~\cite{Fok:2012zk}. 

This result, which fits nicely with the notion of holography, means that a future discovery of particle $X$ is not necessarily a signature of extra-dimensions nor the door to the realm of low-energy quantum gravity.

In my opinion, there are two facts which could suggest that $X$ is not a signature of extra-dimensions, or that these extra-dimensions are not as plain vanilla as we have viewed them so far. 

The first concern regards the effects of quantum gravity, as they are regulated by the same scale $\Lambda$ which determines the couplings of the KK-graviton to SM particles. With $c_{g,\gamma} \lesssim {\cal O}(1)$, $\Lambda$ is about 3-5 TeV and one expects black holes around the same scale, likely quantum black holes~\cite{Meade:2007sz}. Current limits from Run1 LHC8 on these states~\cite{Chatrchyan:2013xva} are right in this region of multi-TeV, which is both a promising prospect for Run2 studies, and a possible way to falsify the KK-graviton hypothesis. 

The other issue with the KK-graviton interpretation is the relation between the mass of a spin-two resonance and KK-vectors in extra-dimensions $V$. It is well known that in AdS warped extra-dimensions, the context used by CMS to interpret $X$, there is a gap between the spin-one and -two KK excitations, $m_X = 1.5 \, m_V$. One would then expect a spin-one resonances below the mass of $X$ at about 500 GeVs, which somehow evaded detection. 

Hiding such a light vector resonance is a non-trivial task, as the KK excitations of SM gauge bosons induce anomalous couplings~\cite{Hirn:2008tc} and can be searched for directly in the $Z'$ and $W'$ type of searches~\cite{}. In the following I am going to assume that, wherever these vector resonances are, they are above the mass of $X$. 

To wit, flipping the gap between $X$ and $V$ is quite challenging. Let us start by trying to remove the vector resonances, localizing the gauge fields within the extra-dimension. Universality of gauge couplings indicates that the fermions are then either localized on the same position or have flat profiles. In either case, exchange of the $X$ particle would lead to sizeable four-fermion interactions~\cite{Rizzo:1999br,Davoudiasl:1999tf,Davoudiasl:2000wi}.

Instead of localizing bulk gauge fields, one could explore the gap between the vector and tensor KK modes in non-trivial geometries for the extra-dimensions. For example, let us consider factorizable metrics of the form
\bea
d s^2 = w(z)^2 \, (\eta_{\mu\nu} d x^\mu d x^\nu - \sum_{i,j} d z_i d z_j ) \ . 
\eea  
Following the techniques in ~\cite{Hirn:2007bb}, one can obtain a sum rule involving the masses of KK towers as a function of the spin~\cite{ross}, which have the typical form 
\bea
\frac{1}{M_{s}^2} \approx \int d z \int^z d y \left(\frac{w(z)}{w(y)}\right)^{2s-1} \ . 
\eea 
As the warping $w$ is positive and monotonically decreasing, one typically reaches the conclusion that even for non-AdS, yet factorizable, metrics
\bea
m_{V} < m_X \ .
\eea
 
Obviously, the extra-dimensional bag of tricks is richer than brane localization or warping. For example, one can try to revert this splitting using localized kinetic or mass terms~\cite{Carena:2002me,Cui:2009dv,Kogan:2001wp,Batell:2005wa}. Finding such configurations would be very interesting.

\section{A glueball of new strong interactions \label{glueball}}

In the previous section we have seen that the interpretation of $X$ as a signature of extra-dimensions would need some model-building efforts to be consistent with the absence of excesses in searches for black holes and spin-one resonances. Whereas one could argue that the modelling of quantum black holes suffers from our lack of understanding of a full theory of quantum gravity, hiding a $Z'$ or $W'$ at 500 GeV seems rather more daunting. 

Focusing then on the issue of the splitting between the vector and tensor resonances, let us discuss it in the context of a strongly coupled theory. Assume there is a new strongly coupled sector at the TeV scale, and $X$ is the first signature of this new sector we have discovered, a spin-two glueball akin to $f_2$ resonance in QCD.  Should we expect a different ordering of spin-one and spin-two resonances in this theory? At this point, some readers may jump to say: {\it "Obviously, the answer is no. Strongly interacting sectors must have a gravitational dual"}. Therefore, the statement that simple gravitational set-ups exhibit vector resonances lighter than tensors surely means the 4D dual must follow the same trend.

That is indeed the case in most known theories, including QCD. In QCD, the lightest spin-two resonance $f_2$ is heavier than the $\rho$ meson, the lightest spin-one resonance. One  can qualitatively understand this fact using the techniques developed by Jaffe, Johnson and Ryzak~\cite{Jaffe:1985qp}.  In their work, they identified the lowest-lying states of a strongly coupled sector using colourless, local operators and classifying them by their canonical dimension. 

This conjecture, although simplistic and full of caveats, does connect with ideas of constituent models, where operators of higher dimension require more excitations. Nevertheless, note that their arguments, outlined below, do not account for the effect in the spectrum of other interactions besides the strong sector, and for the possibility of overlapping states.

Using this dimensional analysis, one can identify the quantum numbers of the lightest mesons in the theory, namely spin and CP, or $J^{CP}$, by decomposing each operator into irreducible representations of angular momentum, parity and charge. Further information can be obtained by using a mean field mode analysis. Generically, one gets

\begin{itemize}
  \item In theories involving massless quarks as constituents, the simplest (lowest-dimension) interpolating fields are dimension-three terms of the form $\bar q \Gamma q$, where $\Gamma$ is any Dirac matrix leading to a colour singlet combination. These operators lead to states with quantum numbers
  \bea
  J^{PC}= 0^{-+} \textrm{  and } 1^{--}
  \eea
  \item On the other hand, the gauge interactions themselves can produce mesons, or glueballs. In this case, the lowest-dimension interpolating operator is dimension-four, $Tr \, G^{\mu \nu} G^{\alpha \beta}$, leading to states of quantum numbers
  \bea
  J^{PC}= 0^{++} \textrm{  and } 2^{++} \ .
  \eea
\end{itemize} 

Let us leave the scalar mesons aside for a moment. In QCD-like theories (theories with chiral symmetries), one would then expect the vector resonances to be lighter than the tensors, as their interpolating operators are of lower dimension (dimension-three vs dimension-four). This is indeed the case in QCD, and lattice simulations of QCD-like theories do seem to fulfill this expectation. 

Coming back to the scalar mesons, these may not appear in the spectrum as narrow resonances and hence evade detection. Such a mechanism has been proposed for the case of the $\sigma$ meson in QCD.  

\vspace{.2cm}
Summing up this discussion, the identification of the $X$ resonance with a spin-two glueball and the absence of a lighter spin-one meson seem to indicate a specific type of strongly coupled sector, namely

\begin{center}
{\it $X$ is the lowest-lying resonance of a confining gauge theory, with no constituent massless fermions.}
\end{center}

\vspace{.2cm}

Based on the statement above, one should look for new scalar resonances (possibly broader) and a nearby heavier spin-two excitation~\cite{Jaffe:1985qp}, which may likely have decays via the lighter $X$.
 
\section{The connection with the Higgs \label{Higgs}}
  
If $X$ is a bound state of a new strong force, we are faced with the question: "Who ordered this sector to confine at around the electroweak scale?" In other words, can the confinement dynamics $\Lambda \sim m_X$ be related to electroweak symmetry breaking and hence the Higgs?

Were the Higgs related to the dynamics at $\Lambda$, one would then suspect the Higgs to be a pseudo-Goldstone boson. Therefore, this new sector should exhibit a global symmetry broken by the new strong sector, leading to the Higgs boson at lower energies.  

At first sight, this does not look like a promising speculation. As we mentioned before, our candidate theory should not have light constituent fermions, otherwise spin-one bound states would dominate the phenomenology. 

Moreover, accommodating a pattern of breaking leading to at least four Goldstone bosons and some amount of custodial symmetry seems like a tall order for a theory without chiral symmetries in the usual (QCD-like) sense.  

Despite all these difficulties, it seems that one can devise theories with massive fermions and global symmetries which, after confinement at $\Lambda_c$, produce a set of Goldstones and no light fermion bound states. Examples of such models are under investigation~\cite{Jack}. 

Finally, it would be interesting to investigate this pattern of breaking in the context of the richer structure of supersymmetric theories, and of non-supersymmetric holographic duals of the pure glue theory.

\section{Summary}
 
In this note, I have discussed the properties of a possible new spin-two resonance $X$ and how these indicate quite specific types of UV completions. 

For example, were $X$ to be a manifestation of new dimensions of space, it could indicate a non-trivial localization of SM particles in the extra-dimension. In the interpretation of $X$ as a glueball of new strong dynamics, the UV theory would seem to be a (mostly) pure-glue confining theory.   
 
Additionally, it would be terribly unnatural for a resonance to show up at the TeV scale, and have no links to electroweak symmetry breaking. In the text, I have argued that this connection may not be straightforward, and may well require new thinking in terms of extra-dimensional set-ups or strongly coupled theories. In particular, achieving a spectrum of a Composite Higgs with a nearby $X$ and no vector resonances, may need some creative thinking on our side as model-builders. 
 
Needless to say, if the existence of a spin-two $X$ were confirmed, it would signal the urgent need for new non-perturbative techniques and the lattice exploration of non-QCD-like theories. Yet, we would certainly have to wait until more data is collected to proclaim a new era for particle physics, and even more data would be necessary to proclaim the discovery of new dimensions of space-time, or of a new force with its own peculiar dynamics.  
  
\section*{Acknowledgements}
I would like to thank my colleagues Stephan Huber and Daniel Litim for discussions during a train ride. This work is supported by the Science Technology and Facilities Council (STFC) under grant number
ST/J000477/1.

\end{document}